\documentclass[10pt,letterpaper]{article}
\usepackage{amsmath}
\usepackage{opex3}
\usepackage{graphicx}
\usepackage{cite}
\usepackage{setspace}

\begin{document}

\title{High-fidelity, broadband stimulated-Brillouin-scattering-based slow light using fast noise modulation}
\author{Yunhui Zhu,$^{1*}$ Myungjun Lee,$^2$ Mark A. Neifeld$^2$\\ and Daniel J. Gauthier$^1$}
\address{$^1$Dept. of Physics, Duke University and the Fitzpatrick Institute for Photonics, Durham, NC, 27708 USA\\

 $^2$Dept. of Electrical Computer Engineering, University of
 Arizona, Tucson, AZ, 85721 USA
}
 \email{yunhui.zhu@duke.edu}

\begin{abstract}
We demonstrate a 5-GHz-broadband tunable slow-light device based on
stimulated Brillouin scattering in a standard highly-nonlinear
optical fiber pumped by a noise-current-modulated laser beam. The
noise-modulation waveform uses an optimized pseudo-random
distribution of the laser drive voltage to obtain an optimal
flat-topped gain profile, which minimizes the pulse distortion and
maximizes pulse delay for a given pump power. Eye-diagram and
signal-to-noise ratio (SNR) analysis show that this new broadband
slow-light technique significantly increases the fidelity of a
delayed data sequence, while maintaining the delay performance. A
fractional delay of 0.81 with a SNR of 5.2 is achieved at the pump
power of 350 mW using a 2-km-long highly nonlinear fiber with the
fast noise-modulation method, demonstrating a 50\% increase in
eye-opening and a 36\% increase in SNR compared to a previous
slow-modulation method.
\end{abstract}

\bibliographystyle{osajnl}

\section{Introduction}

Stimulated-Brillouin-scattering (SBS)-based slow light in room
temperature optical fibers has attracted extensive research interest
over the past few years \cite{boyd2009controlling}. A fiber-based
slow light system can controllably delay optical pulses and can
operate over the entire transparency window of the fiber
\cite{gonzalezherraez2005-optical-tunable-SBS-delay}. However, the
narrow ($\sim$35 MHz) natural linewidth of the SBS resonance (full
width at half magnitude FWHM) in standard single-mode fibers has
limited its application to low-data-rate systems. To solve this
problem, broadband SBS slow-light techniques were developed
\cite{song2007broadbandSBS-25GHz,sakamoto2008distortion-with-comb,gonzalezherraez2006broadbandSBS,zhu2007broadband,yi2007broadbandSBS-35ps,
zadok2006extended,cabreragranado2008Edu's-best-gain,zhu2010competition}.
Herr{\'a}ez \textit{et al.} first used direct current modulation of
a semiconductor laser to broaden its spectrum to $\sim$325 MHz and
hence increased the SBS bandwidth to this value
\cite{gonzalezherraez2006broadbandSBS}. Subsequently, a number of
groups have demonstrated broadband SBS slow-light with bandwidths up
to tens of GHz \cite{zhu2007broadband,yi2007broadbandSBS-35ps,
zadok2006extended,cabreragranado2008Edu's-best-gain,zhu2010competition},
a data rate compatible with modern optical communication systems. In
addition to broadening the spectral linewidth of the SBS resonance,
a judicious choice of the current modulation waveform can be used to
tailor the SBS gain profile, resulting in improved delay performance
for the broadband SBS slow light
systems\cite{yi2007broadbandSBS-35ps,zadok2006extended,cabreragranado2008Edu's-best-gain,zhu2010competition,pant2008distortion-bestgain}.
The optimal gain profile that improves the pulse delay under
constraints of pulse distortion and pump power is a flat-top gain
spectrum with sharp edges
\cite{zadok2006extended,cabreragranado2008Edu's-best-gain,pant2008distortion-bestgain}.
These broadband SBS slow light experiments have extended the
application of SBS slow light to broadband all-optical communication
devices such as data buffering \cite{tucker2005-sl-buffer} and data
packet synchronization\cite{bo2007-sl-synchronization}.

Most previously reported broadband SBS slow light experiments
control the spectral SBS gain profile by direct modulation of the
pump laser using a periodic modulation waveform
\cite{zadok2006extended,cabreragranado2008Edu's-best-gain,zhu2010competition}.
The frequency of the waveform is typically chosen to be in the
sub-MHz range so that detailed features of the waveform can be
reproduced faithfully using an arbitrary waveform generator.
However, such modulation induces low-frequency fluctuations in the
SBS signal, as we discuss below. Previous research that focused on
averaged pulse delay was not affected by these fluctuations because
they were averaged out. Nevertheless, these low-frequency
fluctuations reduces the signal-to-noise ratio (SNR) for a delayed
data sequence and degrades fidelity of the device.

To build an optimal high-fidelity broadband SBS slow light system,
we develop a systematic procedure to generate a broadband
flat-topped SBS gain profile with direct noise current modulation.
Random noise current modulation has been used in previous research
on broadband SBS slow light systems
\cite{gonzalezherraez2006broadbandSBS,zhu2007broadband,yi2007broadbandSBS-35ps}.
However, due to limited control over the spectral profile, these
previous methods generally result in a Gaussian-shaped SBS gain
profile. The frequency-dependent gain of a Gaussian profile causes
pulse distortion for large delays. Although Yi \textit{et al.}
\cite{yi2007broadbandSBS-35ps} have discussed shaping the pump
spectrum by passing a noise waveform through a saturated electronic
amplifier, the control over the SBS gain profile is still limited
and highly sensitive to the detailed saturation characteristics of
the high speed amplifier, which is often hard to characterize. Here,
we present an extension of Yi's method in which the noise
distribution is arbitrarily controlled. Compared to Yi's work, the
method described in this paper is superior because we have complete
control over the noise waveform in a way that is easily generalized
to any DFB laser used as a pump beam in broadband SBS slow light
systems. It will be shown that, by controlling the distribution of
the noise waveform, we are able to tune the shape of the SBS gain
profile and obtain the best flat-topped profile that optimizes the
slow light delay and reduces distortion. We also find that using a
noise modulation function with a sampling rate $\sim$400 MHz (fast
compared to the phonon lifetime ($\sim$ 4 ns) in the fiber)
substantially stabilizes the optical signal and improves the data
fidelity of the broadband SBS slow light system compared to previous
slow modulation methods.

The rest of the paper is organized as follows. Section 2 briefly
reviews the dynamics of a distributed feedback (DFB) laser under
direct current modulation and describes the procedure to obtain a
flat-topped SBS gain profile with two different (slow and fast)
modulation waveforms. Section 3 describes and compares the delay
performance for a 5-Gb/s return-to-zero (RZ) data sequence using
these two methods and quantifies transmission fidelity by
eye-opening (EO) and signal-to-noise ratio measurements. Finally,
our conclusions are summarized in Sec. 4.

\section{Broadband optimal SBS gain profile design with direct current modulation}

In broadband SBS slow light systems, a spectrally broadened laser
 is used as the pump beam. In our experiment, a modulation
voltage $V(t)$ is added to the DC injection current of the DFB laser
via a bias-T (input impedance = 50 $\Omega$). The DFB laser spectrum
is thereby broadened. Broadening of the laser's spectrum with direct
current modulation has been widely used and a quantitative,
semiempirical model for the instantaneous spectral shift of the DFB
laser output due to direct current modulation $i(t)$ has been
established in \cite{zadok1998spectral}. The spectral shift
$\omega_p(t)$ as a function of time is given by

\begin{equation}\label{chirp}
\omega_p(t)=a_0 i(t)-i(t)\otimes h(t),
\end{equation}
where the first term on the right-hand-side of Eq. (\ref{chirp})
represents the linear adiabatic chirp induced by the almost
instantaneous current-related changes of the equilibrium carrier
density, $a_0$ is a constant coefficient, and the second term
describes the slower thermal chirp, which changes the frequency as a
result of temperature-related changes of the refractive index and
physical length of the cavity. The thermal chirp is characterized by
the convolution of $i(t)$ with the impulse response $h(t)=\sum a_n
e^{-\tau/\tau_n}$, where the different time constants $\tau_n$
correspond to thermal conductivities of different layers in the DFB
laser. Measurements for our DFB laser reveal that the dominant
thermal term has a time constant as short as 7.5 ns
\cite{cabreragranado2008Edu's-best-gain}. As a result, an analysis
of both the thermal and adiabatic chirp is necessary to obtain a
precise design of the laser spectrum.

As has been shown in Ref. \cite{pant2008distortion-bestgain}, the
SBS gain profile that optimizes slow-light performance under various
practical constrains is rectangular-shaped with sharp edges and a
flat top. Such a gain profile produces longer delays and reduces
pulse distortion. This is because the flat gain profile enables
uniform amplification over the different frequency components of the
data stream, minimizing the filtering effect and thereby reducing
pulse distortion \cite{stenner2005distortion-with-2lasers}. The
rectangular-shaped gain profile also improves the delay. Using the
Kramers-Kronig relation, the abrupt-edged gain profile increases the
phase shift, which leads to a larger group index and longer delays
\cite{zadok2006extended}. Because the broadband SBS gain profile
$g(\omega_s)$ (where $\omega_s$ is the signal beam frequency) is
given by the convolution of the pump spectrum with the intrinsic
narrow Lorentzian lineshape \cite{zhu2005-SBS-sl-theory}, a
rectangular-shaped pump laser spectrum with a width much greater
than the Lorentzian linewidth produces the desired optimal broadband
SBS gain profile.

We start the design of the optimal SBS gain profile by only
considering the linear adiabatic term in Eq. (\ref{chirp}). In this
case, the frequency distribution of the DFB laser is the same as
that of the current modulation waveform. This is true when the
characteristic time scale of the modulation is faster than any of
the time constants of the DFB laser. When the thermal chirp is
present, the spectral distribution of the noise must be adjusted
using an iterative method, as described below.

\begin{figure}[t]\centering
  \includegraphics[width=11cm]{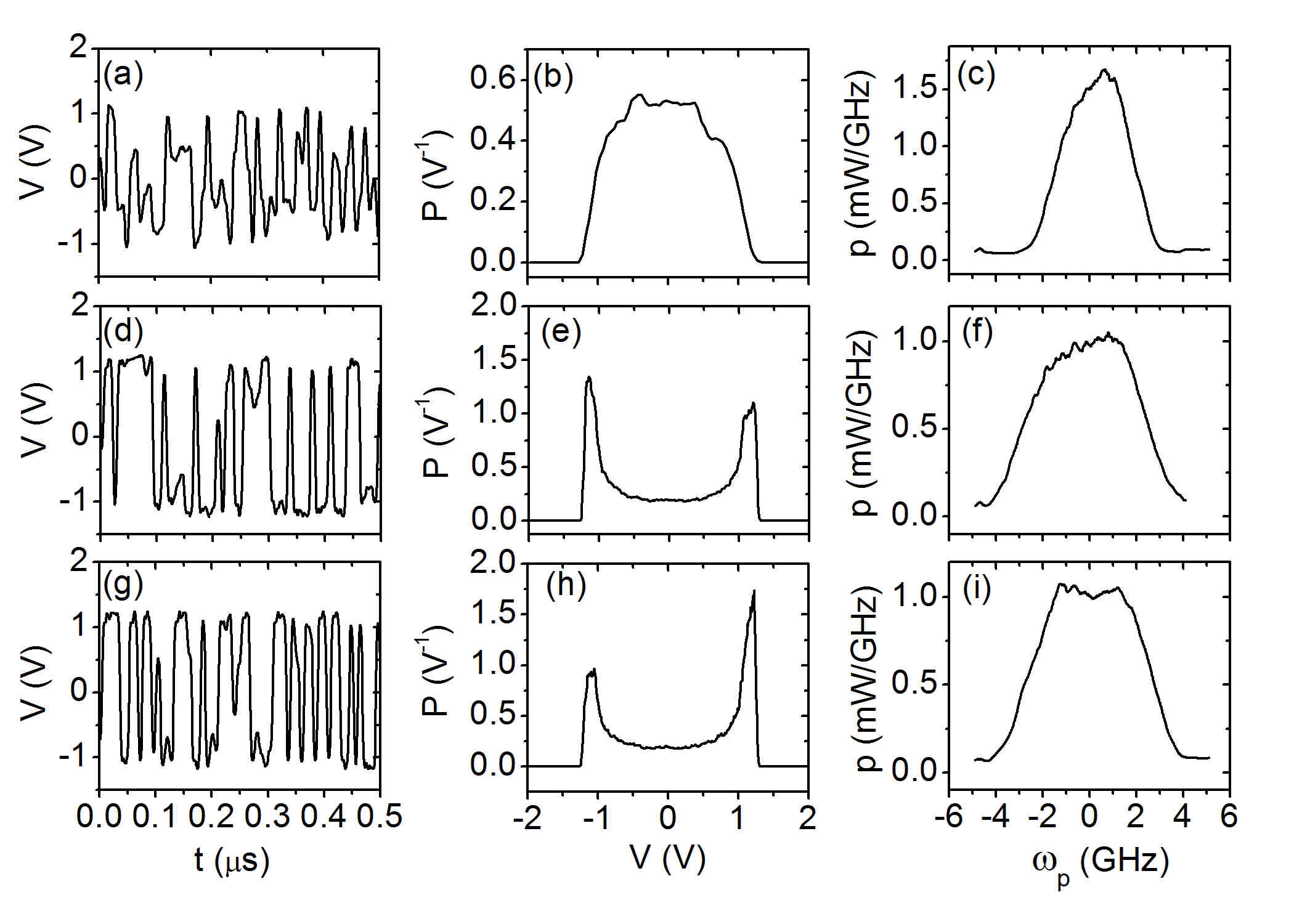}\\
  \caption{Pump spectral distribution optimization procedure for the case of fast noise modulation.
  Modulation voltage waveform $V(t)$ (left column), probability distribution $P$ (bin size = 0.025 V) (middle column) and
  resultant pump beam spectrum $p(\omega_p)$ (right column) are shown for flat-distributed white noise modulation
   $V(t)=2.5$ V$\times f(t)$, where $f(t)$ is a
random variable that is approximate uniformly distributed between
-0.5 and 0.5 (upper row), bi-peak symmetric noise modulation
$V(t)=2.5$ V$\times\tanh [10f(t)]$ (middle row) and optimal noise
modulation $V(t)=2.5$ V$\times\tanh[10(f(t)+0.06)]$ (bottom row).
The DC injection current is 110 mA.}\label{f1}
\end{figure}

 To generate the optimal
rectangular-shaped pump spectrum, we use a noise waveform $V(t)=2.5$
V$\times f(t)$, in which $f(t)$ is a random variable approximately
uniformly distributed between -0.5 and 0.5 (Fig. \ref{f1}(a)). The
sampling time interval is set to 2.5 ns for our arbitrary
wavefunction generator (Tektonix AFG3251). Figure \ref{f1}(b) shows
the probability distribution $P$ of the modulation waveform as a
function of the voltage $V$, which is determined from the histogram
of the waveform. The spectrum of the pump beam $p(\omega_p)$ is
measured by mixing it with a monochromatic reference beam (New Focus
Vortex 6029) on a high-speed detector (New Focus Model 1544b), as
shown in Fig. \ref{f1}(c). We see that the generated pump beam
spectrum shows significant improvement compared to a Gaussian
profile, but is slightly peaked in the center and shows some
asymmetry.

The concentration of the spectrum at the center is due to the
thermal chirp. In particular, the current of the laser is constantly
fluctuating quickly, leading to fluctuations in the temperature
about an equilibrium value. According to Eq. (\ref{chirp}), a step
change in the current $i(t)$ leads to a sudden adiabatic change in
the optical frequency $\omega_p(t)$ followed by thermally induced
exponential decay to a stationary value. The fast noise-modulation
waveform has a rise time $\sim$ 2.5 ns, and has many abrupt changes
that can be considered as instantaneous jumps (Fig.\ref{f1}(a)).
After such an abrupt change, the laser spends some time returning
towards the previous frequency due to the thermal chirp, which
favors frequencies in the middle of the range and causes the
center-concentration effect.

To compensate for this effect, we increase the probability
distribution in the extrema of the noise distribution, using the
function 2.5 V$\times\tanh(bf(t))$. Figure \ref{f1}(d)-(f) show the
waveform $V(t)$, distribution probability $P$, and resultant pump
spectrum $p(\omega_p)$ for $b=10$. We see that the
center-concentration problem in the pump spectrum is solved, but
there is still an asymmetry in the profile, as seen in Fig.
\ref{f1}(f). This asymmetric frequency response is induced by the
nonlinear contribution to the adiabatic chirp (not accounted for in
Eq. (\ref{chirp})) and
 the additional different thermal time constants \cite{cabreragranado2008Edu's-best-gain}.
To solve this problem, an asymmetry is needed in the distribution of
the modulation waveform. We use 2.5 V$\times\tanh[b(f(t)+c)]$, in
which the parameter $c$ controls the asymmetry of the distribution.

The best parameter values for an optimal pump spectrum are obtained
applying an iterative scheme. As we change the parameters in small
steps, the pump spectrum is recorded and compared to an optimal
flat-top spectrum. The error (root mean square deviation RMSD) is
calculated at each step. After a small number of iterations, we find
the combination of parameters that minimizes the error
 using a steepest descent search procedure, which
gives us the optimal values $b=10$ and $c=0.06$.  As shown in Fig.
\ref{f1}(i), modulation with the optimal parameters results in a
good flat-topped spectrum profile with reasonably sharp edges. The
RMSD for this spectral profile is 0.164 mW/GHz, compared to 0.25
mW/GHz for Fig. \ref{f1}(c) and 0.173 mW/GHz for Fig. \ref{f1}(f).

\begin{figure}[b]\centering
  \includegraphics[width=8cm]{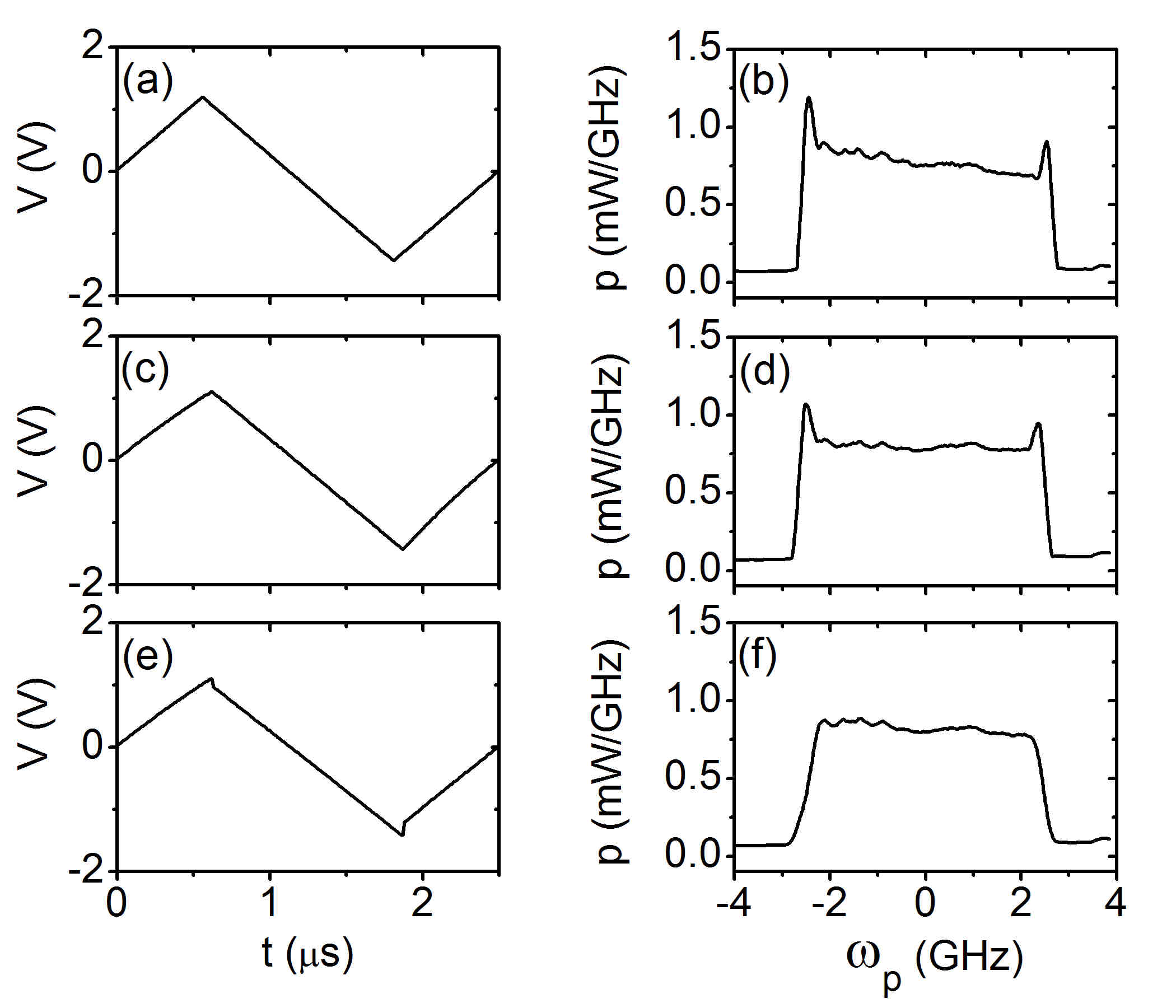}\\
  \caption{Pump spectral distribution optimization procedure for the case of slow modulation. Modulation waveform $V(t)$ (left
  column) and measured pump spectrum profile $p(\omega_p)$ (right column) are shown for
  triangular modulation (upper row), with the addition of a small
quadratic term (middle row), and for the optimum waveform (lower
row). The DC injection current is 110 mA.}\label{f2}
\end{figure}

A similar procedure is used to generate a slow modulation waveform
following Cabrera-Granado's approach
\cite{cabreragranado2008Edu's-best-gain}. We start from a 400-kHz
periodic triangular wavefom and set the amplitude to 2.73 V (Fig.
\ref{f2}(a)). The resultant pump spectrum (Fig. \ref{f2}(b)) shows a
clear asymmetry, which is corrected by introducing a quadratic term
in the triangular waveform (Fig. \ref{f2}(c) and (d)). However, we
still observe peaks at the edge of the spectral  profile induced by
the thermal chirp at the turning points of the waveform. As a result
of the thermal chirp, the instantaneous laser frequency
 spends more time in these regions. These peaks can
be corrected by inducing a current ``jump'' at the turning points,
as shown in Fig. \ref{f2}(e) and (f). The final modulation waveform
is expressed as
\begin{equation}\label{slow}
V(t)=v_{\textrm{max}}\times\left\{
\begin{aligned}
&at^2+(4/T-aT/4)t\quad&\textrm{if}\quad  &t<T/4\\
       &at^2-(4/T+a3T/4)t+2+(2aT^2)/4^2\quad&\textrm{if}\quad  &T/4<t\leq3T/4\\
&at^2+(4/T-a9T/4)t +(5aT^2)/4-4\quad &\textrm{if}\quad &3T/4<t\leq
T,
\end{aligned}\right.
\end{equation}
where $v_{\textrm{max}}=2.73$ V, and $a=-30.4$ $\mu s^{-2}$. The
parameters are optimized using the same error-minimizing iterative
procedure. The RMSD for the optimal spectral profile  is 0.069
mW/GHz (Fig. \ref{f2}(f)), compared to 0.083 mW/GHz for Fig.
\ref{f2}(b) and 0.081 mW/GHz for Fig. \ref{f2}(d).

\begin{figure}[b]\centering
  \includegraphics[width=7.5cm]{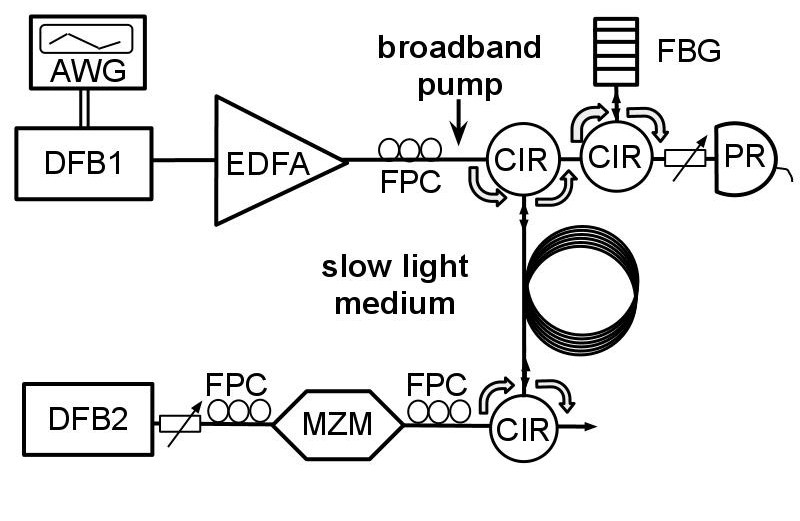}\\
  \caption{Experiment setup. Spectrally broadened pump and signal
  beams counter-propagate in the 2-km-long slow light medium (HNLF, OFS Inc.), where
they interact via the SBS process. The SBS frequency shift in the
NHLF is 9.62 GHz. A fiber Bragg grating (FBG) is used to filter out
the Rayleigh backscattering of the pump beam from the amplified and
delayed signal pulse sequence before detection. AWG: arbitrary
function generator (Tektronix AFG3251), DFB1: 1550-nm DFB laser
diode (Sumitomo Electric, STL4416), EDFA: erbium doped fiber
amplifier (IPG Photonics EAD 1K), DFB2: 1550-n DFB laser diode
(Fitel FOL15DCWC), MZM: Mach-Zehnder Modulator, PR: 12 GHz
photo-receiver (New Focus 1544b), FPC: fiber polarization
controllers, CIR: optical circulator.} \label{f3}
\end{figure}

We then measure the SBS gain profiles produced by the spectral
broadened pump beam using the current modulation waveforms depicted
in Fig. \ref{f1}(g) (the ``fast'' modulation) and Fig. \ref{f2}(e)
(the ``slow'' modulation). The experiment setup is shown in Fig.
\ref{f3}. To independently measure the SBS gain profile, we use a
weak unmodulated monochromatic signal beam (input power $P_{s0}$),
and record the amplified signal beam power $P_s$ at the
photoreceiver as we slowly scan the frequency of the signal beam.
The SBS power gain $G$ is given by
\begin{equation}\label{gain}
G=\ln(P_s/P_{s0}).
\end{equation}
 The SBS power gain $G$ is related
to $g(\omega_s)$ by $G(\omega_s)=g(\omega_s)L_{eff}$, where
$L_{eff}=(1-e^{-\alpha L})/\alpha=1.64$ km is the effective length
of the fiber, $L$ ($=2$ km) is the physical length of the fiber and
$\alpha$ ($=0.9$ dB/km) is the attenuation coefficient of the fiber.

\begin{figure}[h]\centering
  \includegraphics[width=9cm]{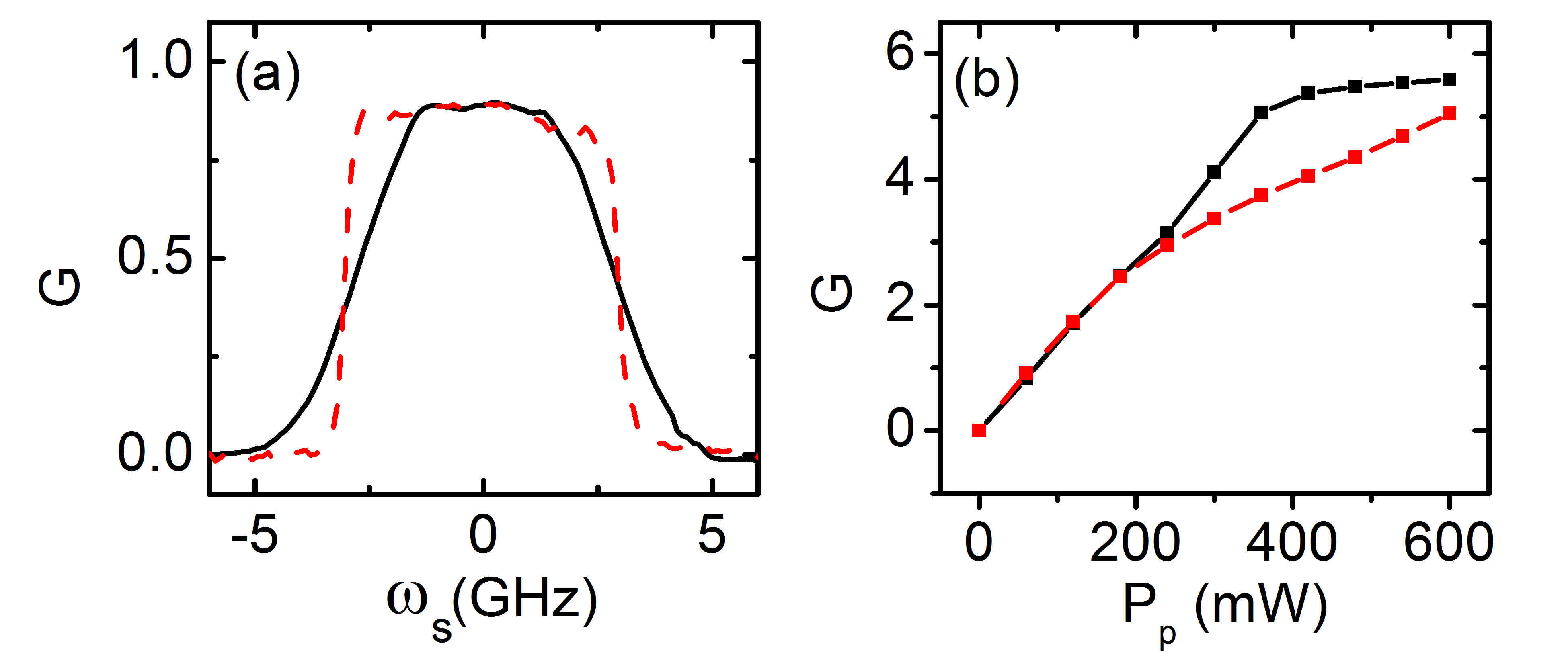}\\
  \caption{(a) SBS gain profiles for fast (solid black line) and
  slow modulations (red dashed line) at $P_p=70$ mW.
  (b) SBS gain saturation for fast and slow modulation methods.
  The black solid line shows the SBS gain $G$ for the fast noise
  modulation, which grows linearly with pump power $P_p$ until
  saturated. The red dashed line shows the SBS gain $G$ for the slow
  modulation, which starts to saturate gradually at a much smaller $P_p$
  compared to the fast modulation method.
  }\label{f4}
\end{figure}

\newpage
Figure \ref{f4}(a) shows the measured SBS gain $G$ profiles for the
fast and slow modulation methods. As discussed previously, the SBS
gain profile is the convolution of the pump spectrum with the
intrinsic narrow Lorentzian lineshape. In our case where the pump
spectrum bandwidth (5 GHz) is much larger than the narrow Lorentzain
linewidth ($\sim$52 MHz in the NHLF), the resultant SBS gain profile
is similar to the pump spectrum, as seen in Fig. \ref{f4}(a). We see
that the SBS gain profile is not as sharp on the edges using the
fast noise-modulation waveform, which is due to the fluctuating
temperature, as discussed above. On the other hand, the slow
triangular-like waveform results in a deterministic value of the
laser temperature at any moment. Therefore, the frequency of the
laser is well-defined at the edge of the modulation waveform.
However, this is not true for the fast noise modulation, where the
temperature is affected by the previous history of the modulation
and thus has wide fluctuations. However, as shown next, the reduced
slope of the edges for the fast noise modulation case does not
substantially affect its slow light performance.

\section{Slow-light performance}

We next compare the delay performance of the fast and slow
modulation methods. First, we use a continuous-wave signal beam
($P_{s0}=48$ $\mu$W) that is tuned to the SBS resonance to measure
the line center SBS gain $G$ at different pump powers. Again, $G$ is
obtained from Eq. (\ref{gain}).  As shown in Fig. \ref{f4}(b), both
modulation formats result in identical linear growth of $G$ with
respect to the pump power $P_p$ when it is low. As $P_p$ increases,
we see that the slow modulation method results in an early
saturation in comparison to the case for the fast modulation
waveform. Saturation takes place when the SBS gain $G$ is large
enough so that a great portion of the power in the pump beam is
transferred into the signal beam, and the exponential amplification
of the signal beam cannot be sustained
\cite{boyd2009nonlinear-optics}.

The early saturation in the slow modulation case is likely due to
fluctuations in $G$. The fluctuation in $G$ is related to the uneven
frequency swept rate and the end effect. In the slow modulation
method, the frequency of the pump beam is slowly swept. During the
modulation period of 2.5 $\mu$s, a monochromatic signal beam is only
intermittently amplified during the short time period when the
pump-probe frequency difference is equal to the SBS frequency shift
within the resonance linewidth. An estimate of the average
interaction time period gives 52 MHz/5 GHz $\times2.5$ $\mu$s$=26$
ns. On this small time scale, sweep rate fluctuations result from
the short thermal constants of the DFB laser can significantly
affect the length of the interaction time period and give rise to
fluctuations in $G$. Moreover, since the SBS amplification process
in the slow modulation method is intermittent, there is an end
effect that induces more fluctuations in $G$. In our experiment
specifically, the frequency of the pump beam as seen by the signal
beam goes through a little less than 8 periods of modulation during
the whole propagation time ($\sim$9.7 $\mu$s) through the 2-km-long
HNLF. Since the number of modulation periods during the propagation
time is not an exact integer, the signal beam can meet the resonant
pump frequency for different times (7 or 8), depending on the
relative time when we measure the waveform during the modulation
period. As a result of both effects, the output signal beam measured
at some particular time is amplified more than others and is more
likely to saturate the gain. This behavior results in the gradual
early saturation seen in Fig. \ref{f4}(b).

In the fast modulation method, on the other hand, a monochromatic
signal is constantly amplified by the frequency-matching component
in the broadband pump beam as it travels through the fiber. The
output signal amplification results from the accumulated SBS
interaction through the whole fiber and has averaged out short-time
fluctuations. Therefore, $G$  is uniform and stable in this case.
The fluctuation in $G$ for the slow modulation method is the source
of the low-frequency fluctuations that degrades the fidelity of a
data waveform, as described next.

To measure the delay and fidelity for a data sequence, we use our
5-GHz broadband SBS slow light system to delay a $2^{12}$ bit-long
return-to-zero (RZ) binary data sequence. This data sequence
contains all $2^8$ 8-bit-long sequences separated by 8-bits 0s
serving as a buffer. In this arrangement, the pattern-dependent
delay is averaged. The use of an RZ signal is more reliable in
situations with pulse broadening effects, but takes twice as much
bandwidth to achieve the same data rate compared to the non-RZ
coding. A data rate of 2.5 Gb/s is used for the signal to match the
SBS slow light bandwidth of 5 GHz (FWHM), where the width of a
single pulse is equal to 200 ps. The data sequence is generated by a
pattern generator (HP70004A) and encoded on the signal beam via the
10-GHz Mach-Zehnder Modulator (MZM). We use a small signal power of
$P_{s0}=12$ $\mu$W and restrict $P_p<500$ mW to avoid SBS gain
saturation. After propagating through the HNLF, the delayed and
amplified signal beam is detected by a 12-GHz photoreceiver and
recorded on an 8-GHz digital sampling oscilloscope (Agilent
DSO80804B). Slow light performance for the fast and slow modulation
methods is evaluated by the well-known fidelity metrics of EO and
SNR based on the eye-diagram of the output signal at various pump
power levels.

\begin{figure}[b]\centering
  \includegraphics[width=11cm]{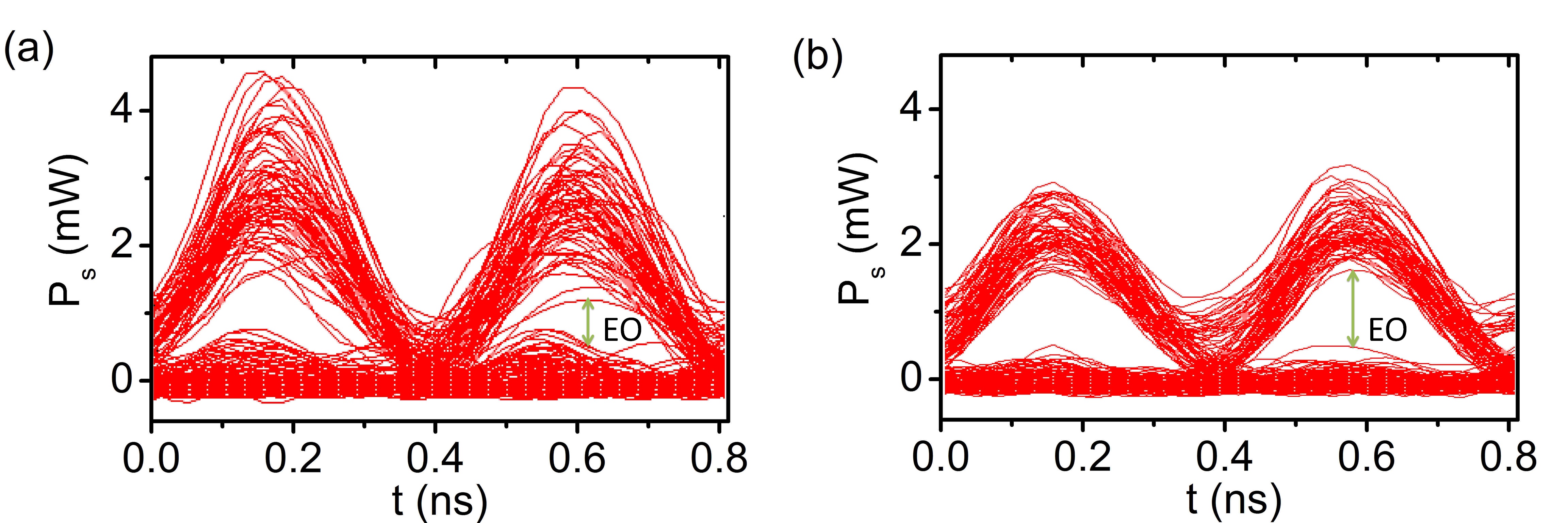}\\
  \caption{Eye diagrams of delayed and amplified data sequences for
   (a) slow and (b) fast  modulation waveforms at $P_p=350$ mW. The arrows in the
   figure show the EO for each case. }\label{f5}
\end{figure}

We first generate the output eye diagram, which is essentially an
overlap of the time domain output traces for a certain number of bit
periods. The EO is measured by the maximum difference between the
minimum value of high level and the maximum value of the low level
in the eye diagram (shown in Fig. \ref{f5}). The pattern delay is
determined by comparing the position of the maximum eye-opening with
and without the pump beam. The SNR at the eye-center is defined as
the ratio of the EO with the quadratic mean of the standard
deviations (noise) of the high and low levels.

\begin{figure}[h]\centering
  \includegraphics[width=12cm]{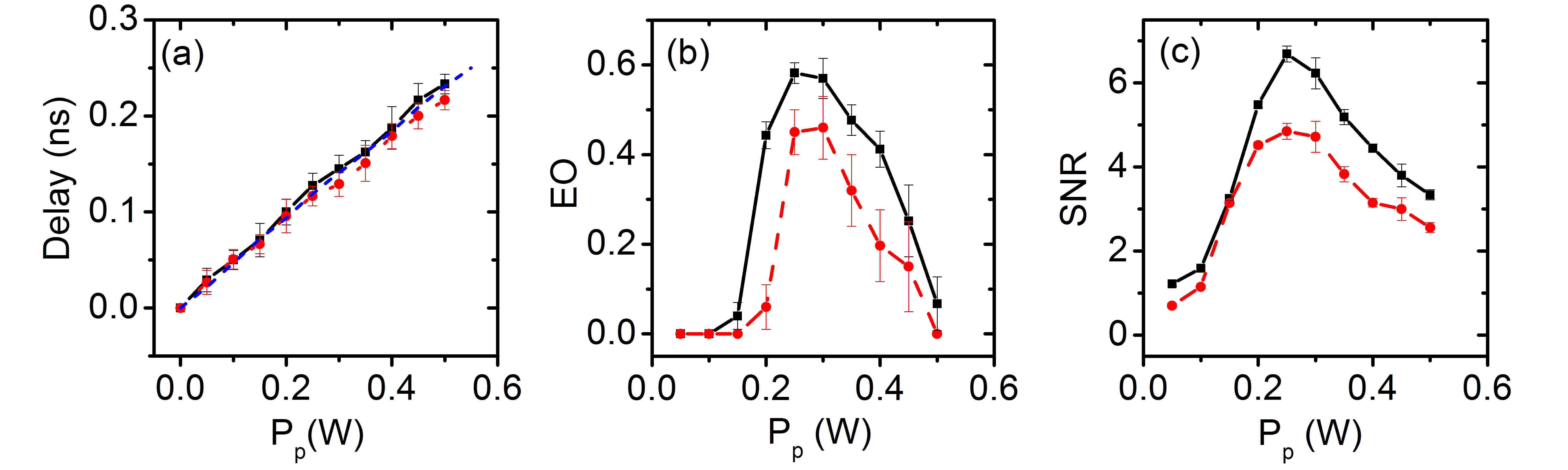}\\
  \caption{(a) Slow light delay and fidelity metrics of (b) EO and (c) SNR
  for fast (solid black line) and slow (dashed red line) modulation waveforms, as functions of
  $P_p$. The theoretically predicted delay for a rectangular-like optimized gain profile
   is also shown in blue short-dash line in (a).}\label{f6}
\end{figure}

Figure \ref{f6}(a) shows the measured pattern delay for both the
slow and fast pump modulation formats as a function of $P_p$. Also
shown are the theoretically predicted delay (blue short-dash line),
assuming a rectangular-like optimized gain
profile\cite{zhu2010competition}. Both modulation formats yield the
same delay
 within the measurement error. The measurements also agree well with the theoretical predictions,
indicating that the deviation from a perfect flat-top rectangular
profile does not substantially degrade the slow light delay.

Figure \ref{f6}(b) shows the EO and Fig. \ref{f6}(c) shows the SNR
as functions of $P_p$. As the output signal beam is amplified with
increasing $P_p$, the signal fidelity first increases as the signal
beam overtakes the detector dark noise, then decreases when the SBS
gain approaches saturation at high pump power, where amplified
spontaneous Brillouin scattering begins to dominate
\cite{boyd1965-noise-both-in-bulk-and-in-fiber}. While both
modulation methods result in similar trends for signal quality at
different pump power levels, the fast noise-modulation method
results in better data fidelity over all pump power levels. In
particular, Fig. \ref{f5} shows an example of the output eye
diagrams for both modulation methods at $P_p=350$ mW. Increased EO
and SNR for the fast noise modulation is clearly demonstrated. A
fractional delay (ratio of the delay with the width of a single
pulse) of 0.81 with a SNR of 5.2 is achieved at $P_p=350$ mW for the
fast modulation method. Compared to the slow modulation method, the
fast modulation method increases the EO by 50\% and SNR by 36\%,
demonstrating significant enhancement of data fidelity with the same
delay.

\section{Conclusion}

We have shown that the signal fidelity is significantly improved in
a broadband SBS slow light system using noise current modulation of
the pump beam spectrum. By controlling the distribution of the
noise-modulation waveform, the SBS gain profile is tailored. We
obtain an optimal flat-topped gain profile using an asymmetric
bi-peak-distributed noise-modulation waveform. Using this new
broadband SBS slow light technique, we significantly improve the
signal fidelity compared to previous low-frequency slow synthesized
waveform modulation methods. Pattern delays up to 1 pulse width is
obtained with high fidelity for RZ data rate of 2.5 Gb/s.

\section*{Acknowledgements}
We gratefully acknowledges the financial support of the DARPA
Defense Sciences Office Slow Light project.
\end{document}